\documentclass{ws-ijmpd}
\usepackage[super,compress]{cite}
\begin{document}

\markboth{J. M. Ladino and E. Larrañaga}
{Motion of a Spinning Particle around an Improved Rotating Black Hole}

\catchline{}{}{}{}{}
%
%%%%%%%%%%%%%%%%%%%%%%%%%%%%%%%%%%%%%%%%%%%%%%%%%%%%%%%%%%%%%%%%%%%%

\title{MOTION OF A SPINNING PARTICLE AROUND AN IMPROVED ROTATING BLACK HOLE}

\author{JOSE MIGUEL LADINO}

\address{Universidad Nacional de Colombia. Sede Bogotá. Facultad de Ciencias.\\ Observatorio Astronómico Nacional. Ciudad Universitaria. Bogota, Colombia.\\
jmladinom@unal.edu.co}

\author{EDUARD LARRAÑAGA}

\address{Universidad Nacional de Colombia. Sede Bogotá. Facultad de Ciencias.\\ Observatorio Astronómico Nacional. Ciudad Universitaria. Bogota, Colombia.\\
ealarranga@unal.edu.co}

\maketitle

\begin{history}
\received{Day May 2022}
\revised{Day Month 2022}
\end{history}

\begin{abstract}
Using the Mathisson–Papapetrou–Dixon equations together with
the Tulczyjew spin-supplementary condition, we study the circular orbits of a spinning test particle moving in the equatorial plane of a quantum improved Rotating Black Hole (RBH) spacetime. This background metric incorporates a position-dependent gravitational constant $G(r)$ and its behavior determines the quantum corrections to the properties of the Innermost Stable Circular Orbit (ISCO). The obtained results show that the radius of the event horizon as well as the radius of the ISCO for the quantum improved RBH are smaller than those of the Schwarzschild or Kerr solutions.
\end{abstract}


\ccode{PACS numbers:04.70.Dy, 04.70.Bw, 11.25.-w}

%\tableofcontents

\section{Introduction}

Black hole solutions are one of the most striking and studied results of General Relativity (GR) and it is well known that one of the phenomena of greatest interest in the analysis of these compact objects is the behavior of a test particle orbiting around them. In fact, since it is known that the characteristics of the geodesics depend on the geometry of spacetime, a focus has been made on the properties that correct them. For example, it has been verified that properties such as the spin or the electric charge of the moving particle affect these orbits \cite{ Zhang2018}. A special case of interest is that of a massive particle orbiting in the Innermost Stable Circular Orbit (ISCO), from which valuable information, e.g. about accretion or radiation spectrum, could be obtained \cite{Zhang2020,Thorne1974}. One of the first works on the properties of the ISCO in a Rotating Black Hole (RBH) was presented by Bardeen, Press and Teukolsky in Ref. \refcite{Bardeen1972},
but they considered only non-spinning test particles. Later, Suzuki and Maeda \cite{Suzuki1998} studied the ISCO numerically using the “pole-dipole” approximation for spinning particles moving in the Schwarzschild and Kerr spacetimes. Then, Jefremov et al. \cite{Jefremov2015} introduced the approximation of small spin, finding some corrections to the ISCO radius and later, the ISCO properties for spinning particles in Kerr-Newman spacetimes has been studied in many references \cite{Zhang2018,Zhang2019,Lukes2017, Zhang2016,Pugliese2013} and for spinning test particles orbiting black hole backgrounds from alternative theories in many others  \cite{Zhang2020, Conde2019,Toshmatov2019,Nucamendi2020, Larranaga2020,Liu2020,An2018}.

In this work, we will consider a black hole background from a quantum gravity scenario. This solution was proposed by Bonanno and Reuter \cite{Bonanno2000} as a spherically symmetric black hole metric in the Quantum Einstein Gravity (QEG) approach (more about QEG theory in Refs. \refcite{Lauscher2002,Reuter1998,Reuter2002}). Their main result is a quantum improved static solution that generalizes the Schwarzschild metric by incorporating a running gravitational constant, $ G (k) $, depending on the energy scale $k$, which reduces to Newton’s universal gravitational constant, $ G_0 $, when taking the appropriate limit \cite{Reuter1998,Bonanno2000}. The behavior of the ISCO for non-spinning particles has been studied in the quantum improved Schwarzschild background in Refs.  \refcite{Zuluaga2021,Rayimbaev2020}.

To implement a more realistic perspective, Reuter and Tuiran \cite{Reuter2011} were the first to obtain a quantum improved RBH solution by applying the QEG approach directly. Similar results were presented in Ref. \refcite{Bonanno2000} and by Bambi and Modesto in Ref. \refcite{Bambi2013}.  An alternative approach, using the well known Newman-Janis algorithm \cite{Newman1965}, was presented by Torres \cite{Torres2017}, obtaining a quantum improved RBH from the static and spherically symmetric solution introduced in Ref. \refcite{Bonanno2000}. 
 Here, we will study the ISCO properties for spinning test particles moving in the quantum improved Kerr spacetime introduced in Refs. \refcite{Torres2017,Reuter2011}. In Sec. II, we will present a brief introduction of the quantum improved RBH. In Sec. III, we obtain the equations of motion for a spinning test particle in the quantum improved RBH background from the Mathisson–Papapetrou–Dixon (MPD) equations \cite{Mathison1937, Papapetrou1951, Dixon1970}. We restrict the equations of motion to trajectories in the equatorial plane in Sec. IV and define the effective potential for the spinning test particle to obtain, numerically, the ISCO properties in terms of the spin of the test particle and the parameters involved in the quantum improved RBH background. Finally, in Sec. V we give some conclusions.

\section{The Quantum Improved Rotating Black Hole }
In order to study some quantum effects in the Schwarzschild metric, Bonanno and Reuter \cite{Bonanno2000} introduce the renormalization group improved Schwarzschild solution,
\begin{align}
ds^{2}  = & -f\left(r\right)dt^{2} +  f\left(r\right)^{-1} dr^{2} + r^2d\Omega^2,\label{eq:MetricSchwarzschild}
\end{align}
where $d\Omega^2 \equiv  d\theta^{2}+\sin^2 \theta d\phi^{2} $ and the lapse function $f(r)$ is
\begin{align}
f(r) = 1 - \frac{2MG(r)}{r}.
\end{align}

The function $G(r)$ is the position-dependent gravitational constant 
\begin{align}
G(r) = \frac{G_0 r^3}{r^3+  \omega G_0\left[r+\gamma G_0 M\right]},\label{eq:runningNewtonconstant}
\end{align}
in which the energy scale dependence, $G(k)$, was transformed into a position dependence, $G(r)$, by using the effective average action \cite{Bonanno2000,Torres2017}. In the above expression we use units in which $c=1$, $G_0$ represents Newton’s universal gravitational constant, M is the black hole’s mass while $ \omega $ and $\gamma$ are constant parameters arising from the non-perturbative renormalization group
theory and from an appropriate cutoff identification, respectively \cite{Bonanno2006,Bonanno2009,Bonanno2000,Torres2017,Torres2013-1,Torres2013-2}.

 When $ \omega =0$, the quantum effects are turned off and Schwarzschild spacetime is recovered. The parameters  are estimated as $\gamma=\frac{9}{2}$ and $ \omega =\frac{167}{30 \pi}$ by comparison with the standard perturbative quantization of Einstein’s gravity. However, in this work we shall treat $ \omega $ and $\gamma$ as free parameters, making use of the fact that the properties of the solution do not rely on their precise values as
long as they are strictly positive \cite{Bonanno2006,Bonanno2009,Bonanno2000,Torres2017,Torres2013-1,Torres2013-2}. 

In Ref. \refcite{Torres2017}, Torres applied the Newman-Janis algorithm \cite{Newman1965} to the line element \eqref{eq:MetricSchwarzschild} and obtained a quantum improved RBH solution that, 
in generalized Boyer-Linquist coordinates, is
\begin{align}
ds^{2}  = & -\left(1 - \frac{2MG(r)r}{\Sigma}\right)dt^{2} +  \frac{\Sigma}{\Delta} dr^{2}  \nonumber \\
 & +\Sigma d\theta^{2} -\frac{4MG(r)ar\sin^2 \theta }{\Sigma} dtd \phi \nonumber \\
 &+  \frac{\left( (r^2-a^2)^2 - a^2 \Delta \sin^2 \theta \right) \sin^2\theta}{\Sigma}  d\phi^{2},\label{eq:Metric}
\end{align}
with 
\begin{eqnarray}
\Delta & = & r^2+a^{2} - 2MG(r)r \\[1ex]
\Sigma & = & r^2 +a^{2}\cos^{2}\theta.
\end{eqnarray}

\begin{figure}[!htp]
    \centering
    \includegraphics[width=1\linewidth]{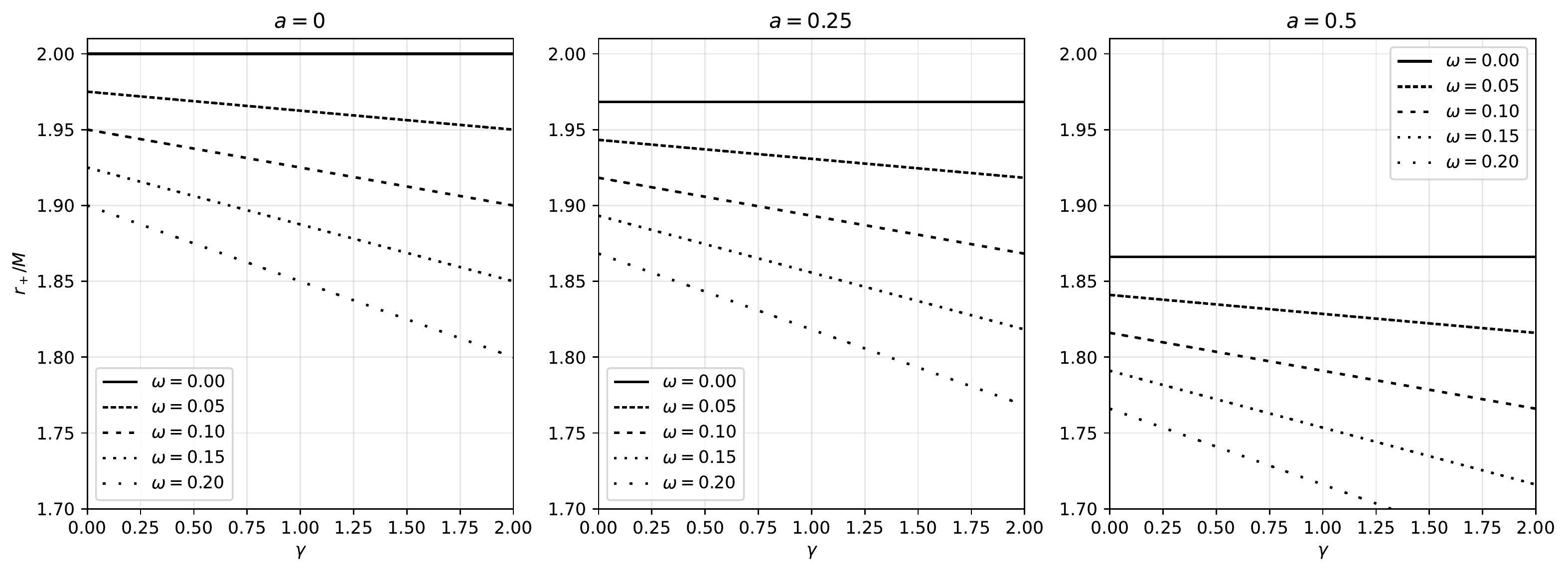}
    \caption{Radius of the event horizon for the quantum improved rotating black hole depending on the parameters $\omega$ and $\gamma$ for different values of the black hole's spin $a$ (using $G_0 = M =1$).}
    \label{fig:EventHorizon}
\end{figure}

This quantum improved RBH solution reduces to Kerr's metric when $ \omega =0$ and reproduces the static solution \eqref{eq:MetricSchwarzschild} when $a=0$. 

The quantum corrected horizons, are defined by the solutions of the equation $\Delta=0$, which corresponds to a fifth-degree polynomial. First, note that the classic limit $ \omega =0$ gives the well-known Kerr horizons,
\begin{equation}
r_{ \pm} = G_0 M \pm \sqrt{G_0^2 M^2-a^2}.\label{eq:Kerrinnerhorizon}
\end{equation}

On the other hand, the roots of the complete equation $\Delta=0$ are, in general, not possible to write in an analytical form. However, it is possible to use a series expansion for the function $G(r)\geq 0$ considering big black hole masses, to find approximate values of the inner and outer horizons \cite{Bonanno2006,Torres2017},
\begin{equation}
r_{-}\approx\frac{1}{2} \sqrt{\gamma   \omega   G_0 + \sqrt{\gamma   \omega   G_0(8 a^2+\gamma   \omega  G_0)}}\label{eq:innerhorizon}
\end{equation}
and
\begin{equation}
r_+\approx G_0 M + \sqrt{G_0^2 M^2 - a^2} - \frac{(2+\gamma)\omega}{4M}.\label{eq:outerhorizon}
\end{equation}

For $a=0$ the resulting horizons are written as 
\begin{align}
r_- \approx & \sqrt{\frac{\gamma \omega   G_0}{2}}\\
r_+\approx & 2G_0 M  - \frac{( 2+\gamma )\omega}{4M},
\end{align}
agreeing with those reported for the improved Schwarzschild metric \cite{Bonanno2006,Torres2017}. 

The behavior of the quantum improved event horizon radius for different values of the spin of the black hole and depending on the parameters $\omega$ and $\gamma$ is shown in Figure \ref{fig:EventHorizon}. Note that the increase of these parameters always produce a decrease in the radius of the horizon.

\section{Equations of Motion for a Spinning Particle}
The equations of motion describing a spinning test particle in the quantum improved RBH background are the MPD equations \cite{Mathison1937, Papapetrou1951, Dixon1970},
\begin{align}
\label{eq:eom1}
\frac{DP^\mu }{D\tau} &= -\frac{1}{2} R^\mu _{ \ \nu \rho \sigma} u^\nu S^{\rho \sigma} \\[2ex]
\label{eq:eom2}
\frac{DS^{\mu \nu} }{D\tau} &= P^\mu u^\nu- P^\nu u^\mu,
\end{align}
where $u^\mu = \frac{dx^\mu}{d\tau}$ is the 4-velocity of the test particle while $P^\mu$ is its 4-momentum, $ R^\mu _{ \ \nu \rho \sigma}$ is the Riemann curvature tensor and $S^{\mu \nu} = - S^{\nu \mu}$ is the spin tensor. The measure of the spin of the test
particle is defined as
\begin{equation}
s^2 = \frac{1}{2} S^{\mu \nu} S_{\mu \nu}.
\label{eq:spin}
\end{equation}

Equations (\ref{eq:eom1}) and (\ref{eq:eom2}) are presented under the “pole–dipole” approximation and may be used to show that due to the spin-curvature force, $u^\mu$ and $P^\mu$ are not, in general, parallel and hence, the trajectory of a spinning test particle wouldn't be a geodesic \cite{Armaza2016,Zhang2020,Zhang2018}. In order to close the system, we will  include the Tulczyjew spin-supplementary condition, 
\begin{equation}
P_\mu S^{\mu \nu} = 0,
\label{eq:Tulczyjew}
\end{equation}
together with the 4-momentum normalization
\begin{equation}
P_\mu P^\mu = - m^2,
\label{eq:momentum}
\end{equation}
being $m$ the proper 
mass of the spinning test particle \cite{Hojman2013,Lukes2014}. However, as noted by many authors \cite{Armaza2016,Zhang2020,Zhang2018},  one must be cautious with the given values for the spin angular momentum $s$ in order to obtain a physically correct 4-velocity. In fact, $u^\mu$ may have a non-timelike character when
$s$ is too large \cite{Suzuki1998} because
\begin{equation}
    u^\mu = \frac{1}{m} \left[ p^\mu + \frac{1}{2m^2\delta} S^{\mu \nu} u^\alpha R_{\nu \alpha \rho \sigma} S^{\rho \sigma} \right]
\end{equation}
with 
\begin{equation}
    \delta = 1 + \frac{1}{4m^2} R_{\alpha \beta \epsilon \lambda} S^{\alpha \beta} S^{\epsilon \lambda}.
\end{equation}

Therefore, following Ref. \refcite{Zhang2018} we will impose the condition given by equation \eqref{eq:superluminal} to restrict those values for the spin that produce an un-physical motion.

\subsection{Tetrad Formalism}
We will introduce a local orthonormal tetrad \cite{Zhang2018,Conde2019,Nucamendi2020,Suzuki1998,Saijo1998} such that
\begin{equation}
g_{\mu \nu} = \eta_{(a)(b)}e^{(a)}_{\mu}e^{(b)}_{\nu} \ \ .
\label{eq:tetraddefinition}
\end{equation}
where $\eta_{(a)(b)}$ represents the usual Minkowski metric tensor and  latin indices label the tetrad. To reproduce the quantum improved RBH spacetime (\ref{eq:Metric}), we use
\begin{align}
\begin{cases}
e^{(0)}_{\mu}dx^{\mu} &= \sqrt{\frac{\Delta}{\Sigma}} \left[  dt - a\sin^{2}\theta d\phi  \right]\\
e^{(1)}_{\mu}dx^{\mu}&= \sqrt{\dfrac{\Sigma}{\Delta}} \, dr  \\
e^{(2)}_{\mu}dx^{\mu} &= \sqrt{\Sigma} \, d\theta \\
e^{(3)}_{\mu}dx^{\mu} &= \dfrac{\sin\theta}{\Sigma} \Big[ -a dt + \left( r^{2} + a^{2} \right) d\phi  \, \Big] . 
\end{cases}
\label{tetrad1}
\end{align}

The inverse tetrad components, defined by the relation
\begin{equation}
e^{\mu}_{(a)} = \eta_{(a)(b)}e^{(b)}_{\ \nu}g^{\mu \nu}  ,
\label{eq:dualtetrad}
\end{equation} 
are
\begin{align}
\begin{cases}
e^{\mu}_{(0)}\partial_{\mu} &= \dfrac{1}{\sqrt{\Delta \Sigma}} \left[ (r^2 + a^2) \partial_t + a \partial_\phi \right] \\ 
e^{\mu}_{(1)} \partial_{\mu} &= \sqrt{\dfrac{\Delta}{\Sigma}} \partial_{r} \\
e^{\mu}_{(2)}\partial_{\mu} &= \dfrac{1}{\sqrt{\Sigma}} \partial_{\theta} \\
e^{\mu}_{(3)}\partial_{\mu} &= \dfrac{1}{\sqrt{\Sigma} \sin \theta} \left[ a \sin^2 \theta \partial_t +  \partial_{\phi} \right].
\end{cases}
\label{eq:tetrad2}
\end{align}

\subsection{Conserved Quantities}
The quantum improved RBH background clearly admits two Killing vectors. 
First, conservation of energy is related to the timelike Killing vector $\xi^{\mu} =\frac{\partial}{\partial t}$ and second,  conservation of the total angular momentum relates to the spacelike Killing vector $ \varphi^{\mu} =  \frac{\partial}{\partial \phi}$ \cite{Zhang2020,Conde2019,Larranaga2020,Toshmatov2019,Nucamendi2020, Suzuki1998,Saijo1998}. The components of these vectors  in the tetrad frame are
\begin{align}
 \xi^{(a)} & = \frac{1}{\sqrt{\Sigma}} \left( \, \sqrt{\Delta} \, , \ 0 \, , \, 0 \, , \, -a\sin \theta \, \right)\,  \\  
 \varphi^{(a)} & = \frac{\sin \theta}{\sqrt{\Sigma}} \left( - \sqrt{\Delta} a \sin \theta \, , \, 0 \, , \, 0 \, , \, (r^2 + a^2)  \right).
\end{align}

The relation between a Killing vector $k$ and the corresponding conserved quantity $C_{k}$ is given by
\begin{equation}
C_{k} = P^{\mu}k_\mu + \dfrac{1}{2}S^{\mu \nu} \nabla_{\mu} k_\nu.
\label{eq:conserved}
\end{equation}

\subsection{Equatorial Motion}
In order to simplify the study of circular motion of a spinning test particle, we will restrict the orbital motion to the equatorial plane. This implies that $\theta = \frac{\pi}{2}$ and therefore, considering that the non-vanishing component of the spin vector is $s^{(2)} = -s$ \cite{Saijo1998,Suzuki1998}, the motion will have two possible situations: spin-aligned or spin-anti-aligned orbits. Furthermore, equatorial motion implies that
the 4-momentum and spin tensor satisfy  \cite{Zhang2020,Saijo1998}
\begin{align}
P^{(2)}=&0 \\
S^{(2) (a)}=&0,
\label{eq:equatorialcondition2}
\end{align}
which, together with equations \eqref{eq:spin} and \eqref{eq:Tulczyjew}, result in the following nonvanishing components
\begin{equation}
\begin{aligned}
\begin{cases}
S^{(0)(1)} &= -sP^{(3)} =- S^{(1)(0)}  \\[0.5ex]
S^{(0)(3)} &= s P^{(1)} = -S^{(3)(0)} \\[0.5ex]
S^{(1)(3)} &= sP^{(0)} = - S^{(3)(1)}.
\end{cases}
\end{aligned}
\label{eq:spincomponents}
\end{equation}

The conserved quantities arising from equation \eqref{eq:conserved} are identified with the energy $C_{\xi}=-E$ and the angular momentum $C_{\varphi}=J$. Using \eqref{eq:spincomponents} these quantities are
\begin{align}
- E=&  -P^{(0)}\xi^{(0)} + P^{(3)}\xi^{(3)} -  sP^{(3)} \nabla_{(0)} \xi_{(1)} +sP^{(1)} \nabla_{(0)} \xi_{(3)} + sP^{(0)} \nabla_{(1)} \xi_{(3)} ,\label{eq:Energy}\\
J = &- \varphi^{(0)}P^{(0)} + \varphi^{(3)}P^{(3)} - sP^{(3)} \nabla_{(0)} \varphi_{(1)}  + sP^{(1)} \nabla_{(0)}\varphi_{(3)} + sP^{(0)} \nabla_{(1)}\varphi_{(3)},
\label{eq:Momentum}
\end{align}
where the non-vanishing components of the covariant derivatives of the Killing vectors are
\begin{align}
 \nabla_{(0)}\xi_{(1)} = &- \dfrac{Mr}{\Sigma^{2}}\left[G(r)r  -  \Sigma G'(r) \right] \notag \\
\nabla_{(0)}\varphi_{(1)} = &-\frac{a}{\Sigma^2} \left[ r\Sigma +MG(r) r^2 - Mr \Sigma G'(r) \right] \notag \\ 
\nabla_{(1)}\varphi_{(3)} =& \frac{\sqrt{\Delta}}{\Sigma} r .
\end{align}

The first derivative $G'(r)$ of the running gravitational constant \eqref{eq:runningNewtonconstant} is presented in equation \eqref{eq:firstderG} in the Appendix.

From equations \eqref{eq:Energy}, \eqref{eq:Momentum} and \eqref{eq:momentum} we obtain the components of the 4-momentum as
\begin{equation}
\begin{aligned}
P^{(0)} = &\dfrac{r}{\sqrt{\Delta}} \dfrac{X}{Z}\, , \hspace{1cm} P^{(1)} = \pm \dfrac{\sqrt{\mathcal{R}}}{\sqrt{\Delta}Z}\, , \\[2ex]
P^{(2)} = &0\, , \hspace{1.85cm} P^{(3)} = r^{2}\dfrac{Y}{Z} \, , \\[2ex]
\end{aligned}
\end{equation}
where we introduce the functions
\begin{equation}
\begin{aligned}
X = &\left[ r^3 + a^2r + as \left( r + M(G(r) - r G'(r))\right) \right] E \\ 
& - \left[ar + Ms(G(r) - r G'(r))  \right] J \\[2ex]
Y = &  J - (a+s)E  \\
Z =& r^3 - Ms^2 \left[ G(r) - r G'(r)\right] \\
\mathcal{R} = &r^2 X^2 - \Delta \left( r^4 Y^2 + m^2 Z^2 \right).
\end{aligned}
\label{eq:AuxFunctions}
\end{equation}

The components of the 4-momentum in the coordinate frame are obtained from the projection $P^{\mu} = e^{\mu}_{(a)} P^{(a)}$, which gives 
\begin{equation}
\begin{cases}
P^0 = & \frac{1}{\Delta Z} \left[ (r^2 + a^2)X + ar \Delta Y \right]  \\
P^1 = & \pm \frac{\sqrt{\mathcal{R}}}{rZ} \\ 
P^2 = & 0  \\
P^3 = & \frac{1}{\Delta Z} \left[ aX + r \Delta Y\right].
\end{cases}
\end{equation}

Using the normalization of the momentum vector \eqref{eq:momentum} and the Tulczyjew spin-supplementary condition \eqref{eq:Tulczyjew}, we obtain the spin tensor components as
\begin{align}
S^{01} = -\dfrac{S^{31}P_{3}}{P_{0}} \ \ , \hspace{1.5cm} S^{03} = \dfrac{S^{31}P_{1}}{P_{0}}.
\label{eq:Spin4-Momentum}
\end{align}

Replacing these expressions in equation \eqref{eq:spin}, gives
\begin{equation}
S^{31} = \dfrac{sP_{0}}{m} \sqrt{\dfrac{g_{22}}{-g}}
\label{eq:Spin4-Momentum2}
\end{equation}
where $g = \textrm{det} ( g_{\mu \nu} ) = -r^4$. Hence, the nonvanishing spin tensor components are
\begin{equation}
\begin{cases}
S^{01} = &-S^{01} = \dfrac{sP_{3}}{mr} \\
S^{03} = &-S^{30} = - \dfrac{sP_{1}}{mr} \\
S^{31} = &-S^{13} = \dfrac{sP_{0}}{mr}.
\end{cases}
\label{eq:spincomponents2}
\end{equation}

\subsection{The Equations of Motion}
Replacing the above results in the MPD equations, \eqref{eq:eom1} and \eqref{eq:eom2}, gives
\begin{align}
P^{0}\dot{r}-P^{1} =& \frac{s}{2mr} g_{3\mu}R^{\mu}_{\ \nu \rho \sigma}\dot{x}^{\nu}S^{\rho \sigma} + \frac{sP_3}{mr^2}\dot{r}  \\
P^{0}\dot{\phi} - P^{3} =& - \frac{s}{2mr}g_{1\mu}R^{\mu}_{\ \nu \rho \sigma}\dot{x}^{\nu}S^{\rho \sigma} -\dfrac{sP_{1}}{mr^2}\dot{r}
\end{align}

By replacing the relevant non-vanishing components of the Riemann tensor (presented in the Appendix), we obtain the following equations for the velocity
\begin{align}
\dot{r} = &P^1\left[ 1 + \frac{s^2}{m^2r^2}g_{11} R_{3003} \right] \left[ P^0  + \frac{s^2}{m^2r^2} \left(R_{3113}P_0 + R_{3101}P_3\right) - \frac{s}{mr^3}P_3 \right]^{-1}
	\label{eq:radialvelocity}\\
\dot{\phi} = &\left[P^3 + \frac{s^2}{m^2 r^2} \left( R_{1001}P_3 + R_{1013}P_0 \right) - \frac{sP_1}{m r^2} \dot{r} \right]  \left[ P^0 - \frac{s^2}{m^2 r^2} \left(R_{1301}P_3 + R_{1313}P_0 \right) \right]^{-1}.
	\label{eq:angularvelocity}
\end{align}

\section{Effective Potential}
Since the radial velocity and the radial component of the momentum are parallel (see equation \eqref{eq:radialvelocity}), the condition for a circular orbits, $\dot{r}=0$, is equivalent to impose  $P^1=0$, which can be written as
\begin{align}
(P^1)^{2} =0= &\frac{m^2}{r^2 Z^2} \left(A e^{2} + B e + C\right)\notag \\
	= &\frac{m^2 A}{r^2 Z^2} \left(e -\frac{-B+\sqrt{B^{2}-4AC}}{2A}\right) \left(e+\frac{B+\sqrt{B^{2}-4AC}}{2A}\right),
	\label{eq:squaremomentum}
\end{align}
where 
\begin{align}
A =& r^2 \left[ K_1^2 - \Delta r^2 (a+s)^2 \right] \\
B =& 2 r^2  j \left[ K_1 K_2 - \Delta r^2 (a+s) \right] \\
C =& r^2 j^2 \left[ K_2^2 - \Delta r^2 \right] - \Delta Z^2\\
K_1 =& r^3 + a^2 r + as \left[ r + M(G(r) - r G'(r))\right]\\
K_2 =& - \left[ ar + Ms(G(r) - r G'(r) )\right].
\end{align}
and we have introduced the quantities $e = \frac{E}{m}$ and $j=\frac{J}{m} = \frac{\ell+s}{m}$ with $\ell$ the orbital angular momentum. The roots obtained from equation \eqref{eq:squaremomentum} define the effective potential as 
\begin{equation}
V_{\textrm{eff}} (r) =  \frac{-B + \sqrt{B^2 - 4 A C}}{2 A}.
\label{eq:V-effective}
\end{equation}

\begin{figure}[!htp]
    \centering
    \includegraphics[height=7cm, width=8cm]{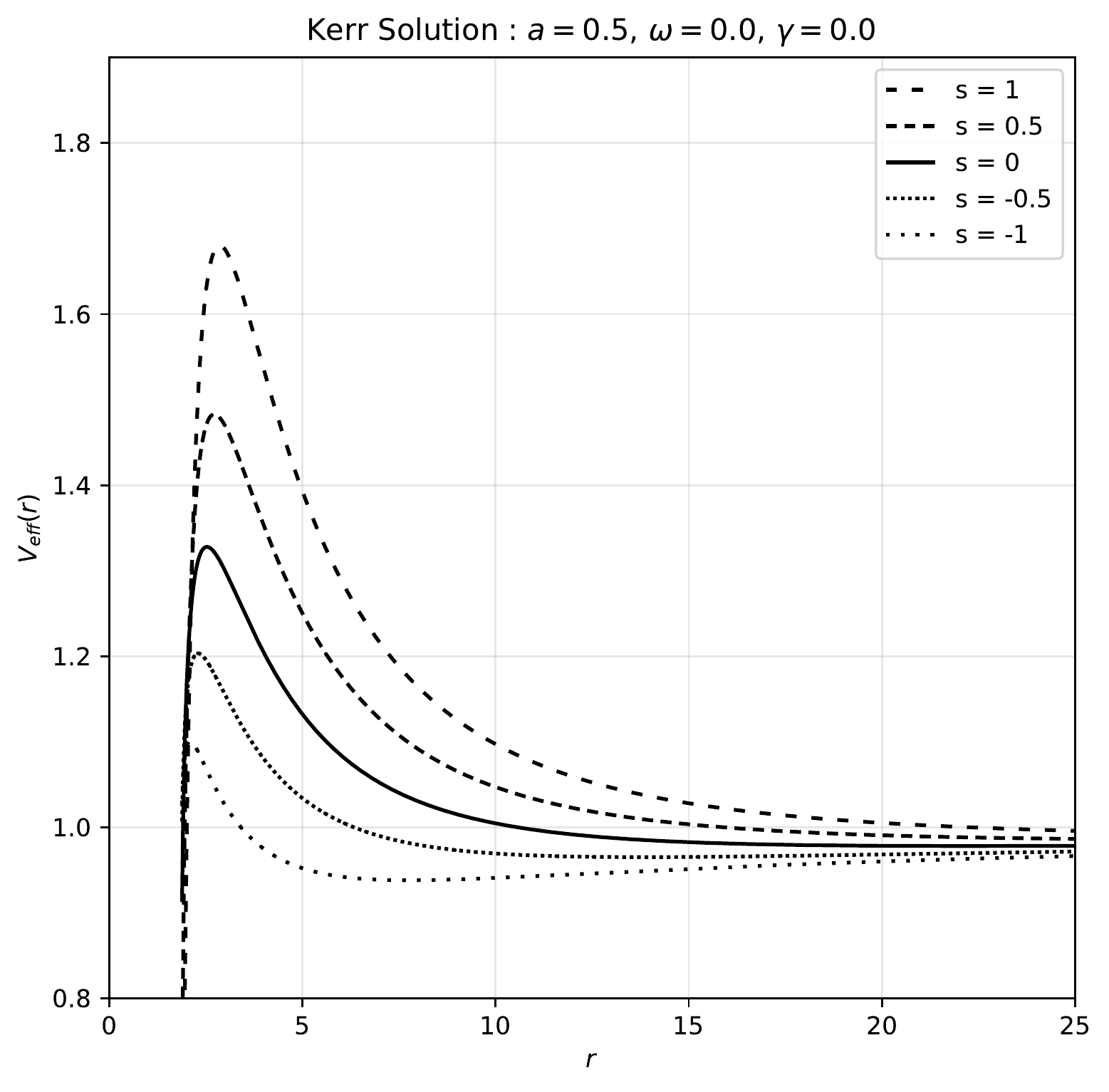}
    \caption{Effective potential for spinning test particles with orbital angular momentum $\ell = 5.0$, moving in Kerr's background with $a=0.5$ (using $G_0 = M =1$ and measuring $s$ in units of $m$).}
    \label{fig:effPotentialKerr}
\end{figure}

It is noteworthy that we consider only the positive square root because the motion of
a spinning particle should be future directed \cite{Saijo1998}. The effective potential in the classical limit is obtained by ignoring the quantum corrections, $\omega  =0$. Figure \ref{fig:effPotentialKerr} shows the typical behavior of effective potential as function of the radial coordinate for different values of spin. Positive values of $s$ describe the case of spin aligned with the orbital momentum while negative values represent the anti-aligned case \cite{Saijo1998}. In this figure, as well as in all the following plots, we use $m=1$. The orbital angular momentum value is chosen as $\ell= 5.$ and the black hole spin as $a= 0.5$.

%Obviously, as we can see in Fig. \ref{fig:effPotentialKerr} and as has already been demonstrated before, the effective potential for a spinning test particle around an RBH is being affected by its spin \cite{Saijo1998,Suzuki1998,Jefremov2015,Zhang2018,Zhang2019,Lukes2017, Zhang2016,Pugliese2013}.

\begin{figure}[!htp]
    \centering
    \includegraphics[width=1\linewidth]{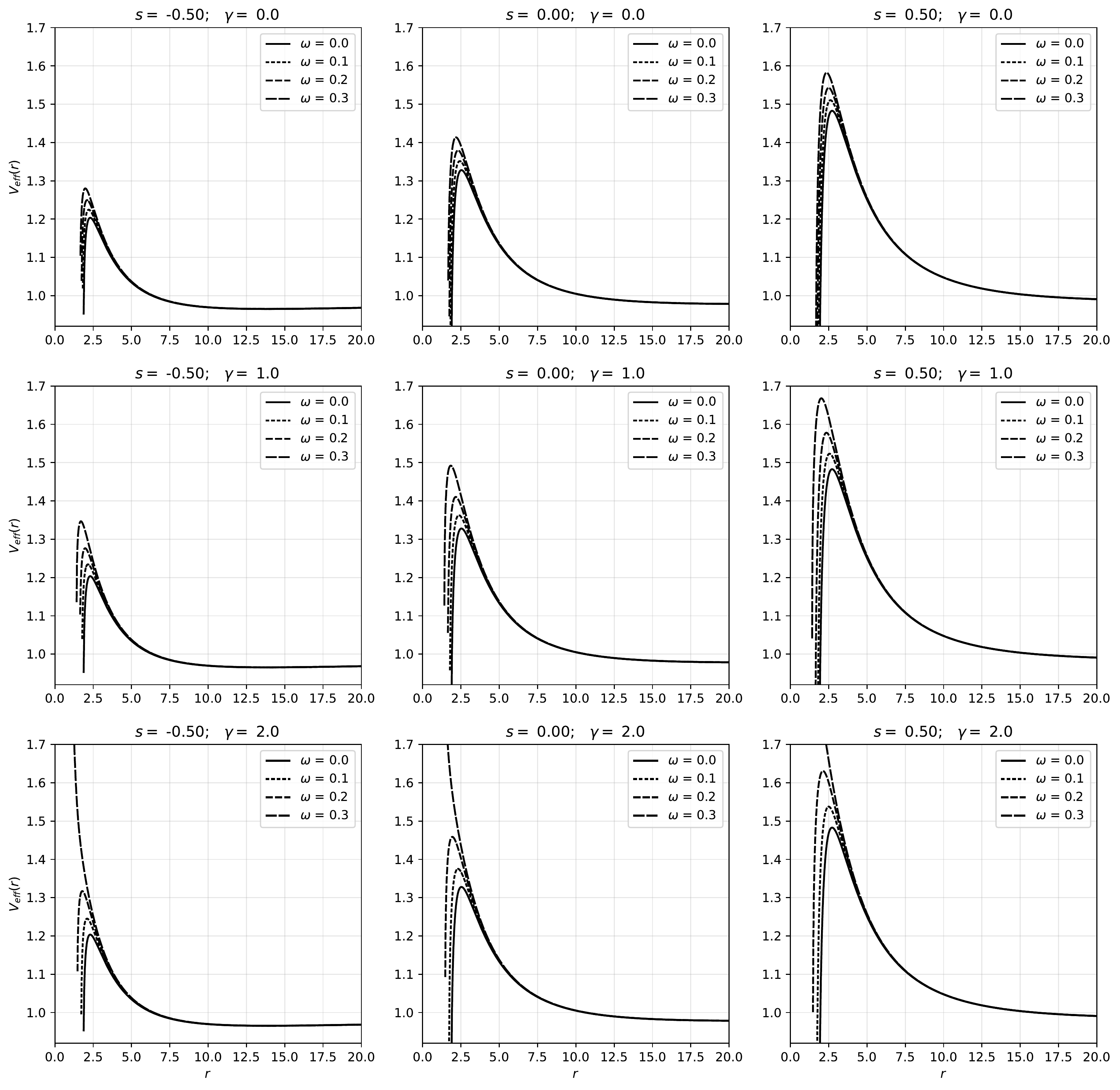}
    \caption{Effective potential for a spinning test particle moving in the quantum improved RBH geometry with orbital angular momentum $\ell = 5$ and $a=0.5$ (using $G_0 = M =1$ and measuring $s$ in units of $m$).}
    \label{fig:effPotentials}
\end{figure}

Figure \ref{fig:effPotentials} shows some of the possible cases of the effective potential for a spinning particle moving in the quantum improved RBH background with different elections of the parameters $\gamma$ and $ \omega $. It is evident that these quantum parameters affect the effective potential of the spinning test particle and, particularly, the location of its minimum and maximum, which define the circular orbits.

\section{ISCO of a Spinning test Particle in a quantum improved RBH background}
The ISCO orbit is defined by the relations 
\begin{align}
\frac{dV_{\textrm{eff}}}{dr} = 0, \ \  \hspace{1.5cm} \frac{d^2 V_{\textrm{eff}}}{dr^2} = 0 .
\end{align}

By numerically solving these conditions, we obtain the ISCO parameters $r_{ISCO}$, $e_{ISCO}$ and $\ell_{ISCO}$ for different values of $s$, $\gamma$ and $ \omega $. Additionally, the numerical solution is checked to satisfy the timelike constraint for the velocity,
\begin{equation}
    \frac{u^2}{(u^0)^2} = \frac{u^\mu u_\mu}{(u^0)^2} = g_{00} + g_{11}\dot{r}^2+ 2 g_{03} \dot{\phi}  + g_{33} \dot{\phi}^2  < 0,
    \label{eq:superluminal}
\end{equation}
ensuring that the motion of the spinning test particle is physical \cite{Zhang2018}.

\subsection{Schwarzschild and Kerr Backgrounds}

\begin{figure}[!htp]
    \centering
    \includegraphics[width=1\linewidth]{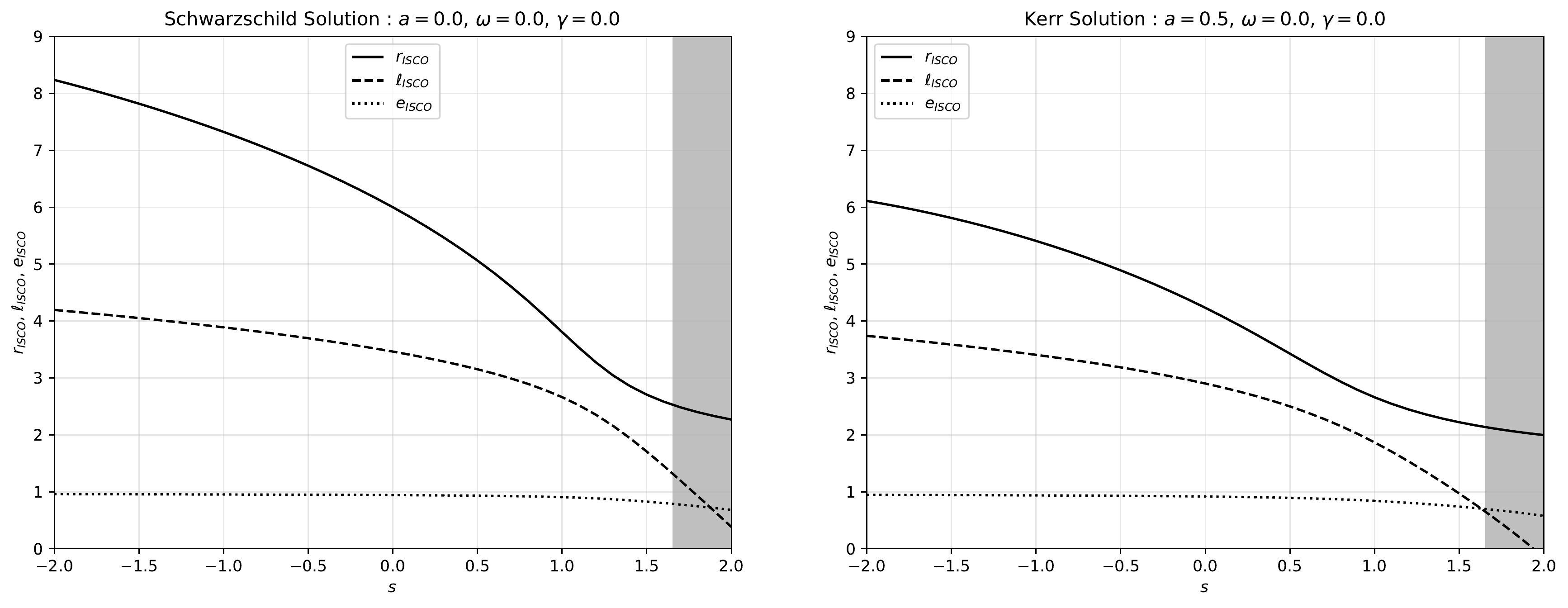}
    \caption{ISCO parameters as function of the spin $s$ for a test particle moving in the Schwarzschild and Kerr backgrounds. The shaded regions correspond to non-physical motion (using $G_0 = M =1$ and measuring $s$ in units of $m$).}
    \label{fig:Sch-Kerr-ISCO}
\end{figure}

The first case that we analyze is the particular limit of classical black holes ($ \omega  =0$), specifically both the Schwarzschild ($a=0$) and the Kerr ($a \neq 0$) solutions. Figure \ref{fig:Sch-Kerr-ISCO}, shows the values of the physical properties for the motion of the test particle at the ISCO of these backgrounds. It should be noted that physical motion of the particle is possible in the non-shaded shaded region. That is to say, in the shaded region the velocity of the particle is spacelike and hence, unphysical \cite{Zhang2018}. These results agree with those presented in Refs. \refcite{Zhang2018,Jefremov2015}. For example, the radius $r_{ISCO}$ for non-spinning particles ($s = 0$) corresponds to the well-known radius behavior for the Schwarzschild solution $r_{ISCO}=6M$ and for the Kerr case, in which we have $2M \le r_{ISCO}\leq 9M$ for $0\le a \le 1$. Additionally, the radius of the ISCO for the spinning test particle in the parallel case, $s > 0$, is always smaller than that of the non-spinning test particle in the Schwarzschild and Kerr black holes backgrounds. In the anti-parallel case, $s < 0$, the radius of the ISCO is always larger than that of the non-spinning test particle.

\subsection{Quantum Improved Black Holes}

\begin{figure}[!htp]
    \centering
    \includegraphics[width=1\linewidth]{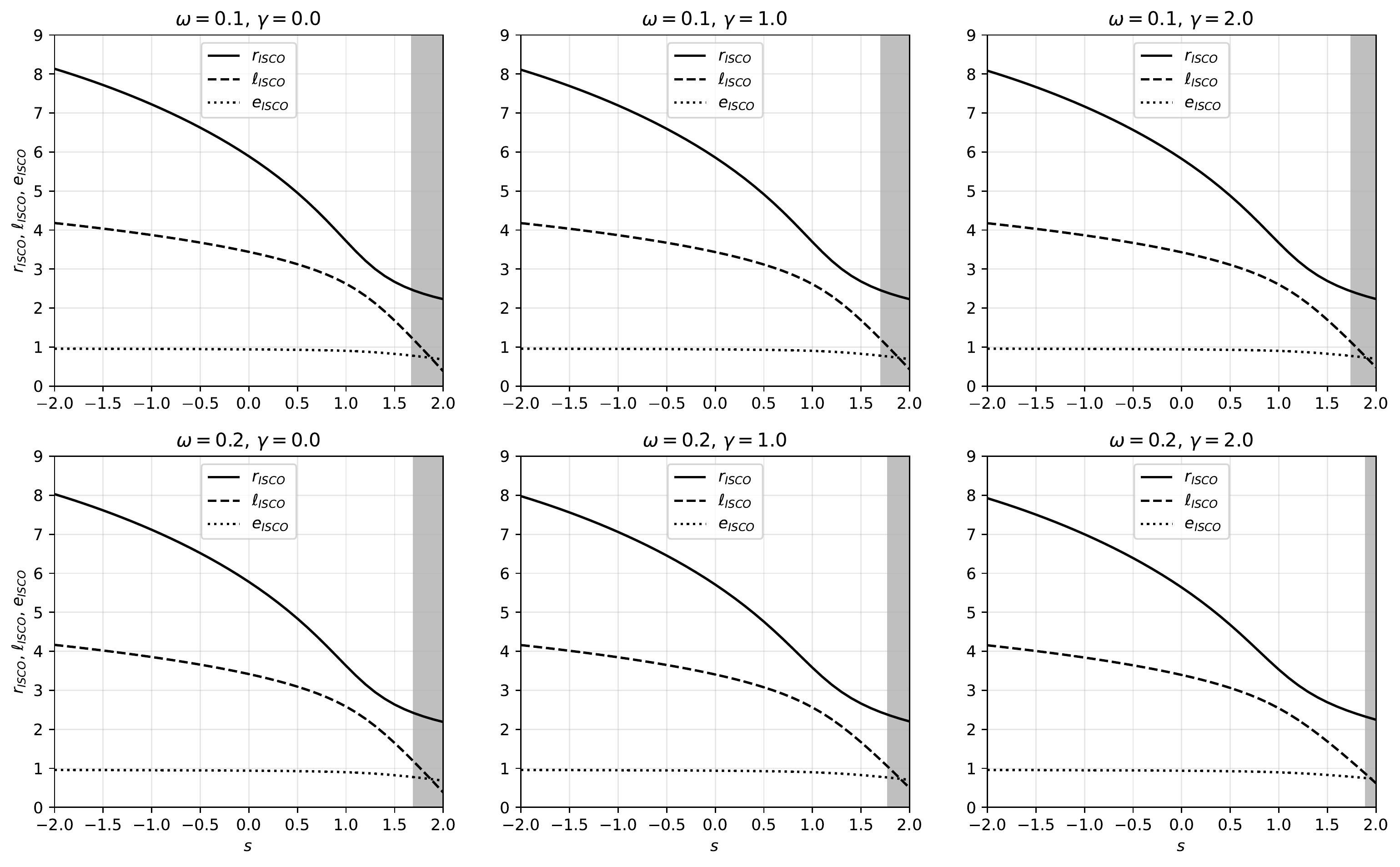}
    \caption{ISCO parameters as function of the spin $s$ for a test particle moving in the quantum improved Schwarzschild black hole background. The shaded regions correspond to superluminal motion (using $G_0 = M =1$).}
    \label{fig:Schquantum-ISCO}
\end{figure}

In order to estimate the effects of the quantum corrections in a non-rotating background, we proceed to vary the corresponding quantum parameters $\omega$ and $\gamma$ as shown in Figure \ref{fig:Schquantum-ISCO}. The general behavior of the ISCO parameters is the same, all quantities continue to decrease under an increase of the spin value, being lower in the parallel case and higher in the non-parallel case. In particular, looking at the case of non-spinning particles, it is possible to notice that the classical ISCO radius decreases, so that $ r_{ISCO}<6M $ due to the quantum effect. For example, using $\omega=0.2$ and $\gamma=2.0$ the ISCO for non-spinning particles have a radius of $ r_ {ISCO} \approx 5.6377 M $. 

A similar behavior is seen for the angular momentum $\ell_{ISCO}$ and for the energy $e_{ISCO}$. However, the variation is more evident for the radius of the ISCO.

\begin{figure}[!htp]
    \centering
    \includegraphics[width=1\linewidth]{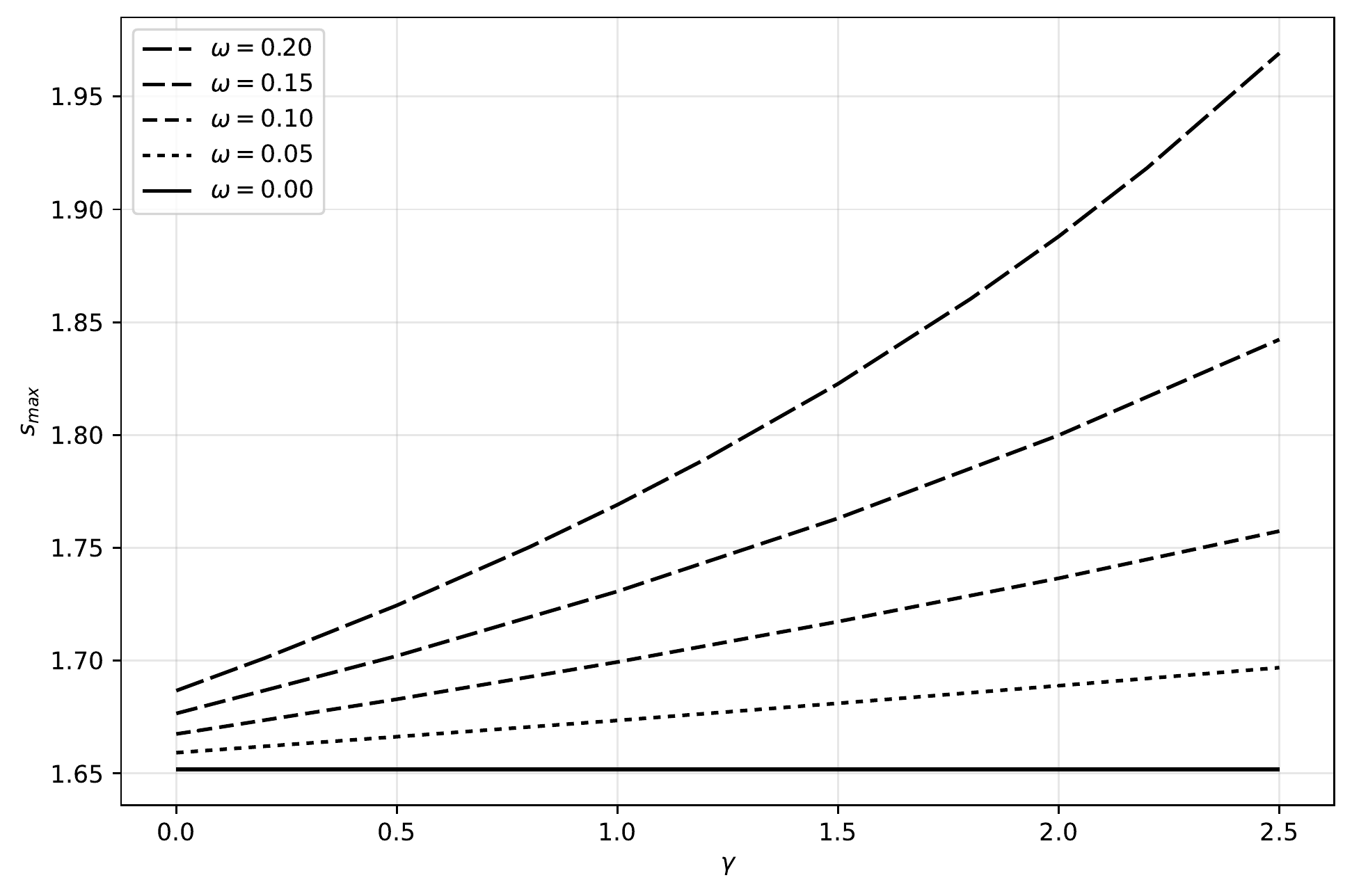}
    \caption{Maximum spin allowed by the velocity constrain as function of the parameters $\omega$ and $\gamma$ for the improved Schwarzschild metric (using $G_0 = M =1$ and measuring $s$ in units of $m$).}
    \label{fig:Schmaxspin}
\end{figure}

The value of the maximum allowed spin, $s_{max}$, delimiting the shaded regions in Figure \ref{fig:Schquantum-ISCO} and defining the regions of the non-physical motion, depends on the parameters $\gamma$ and $ \omega$ through equation \eqref{eq:superluminal}.  Figure \ref{fig:Schmaxspin} shows this dependence, giving the maximum allowed spin in the quantum improved non-rotating spacetime. The results suggest that  higher  values of the quantum parameters imply a larger value of the maximum spin. For example, by taking $\omega=0.20$ and $\gamma=2.5$ we have a maximum spin of  $s_{max} \approx 1.8423$, while the classic case, $ \omega  =0$, has $s_{max} \approx 1.6518$.

\begin{figure}[!htp]
    \centering
    \includegraphics[width=1\linewidth]{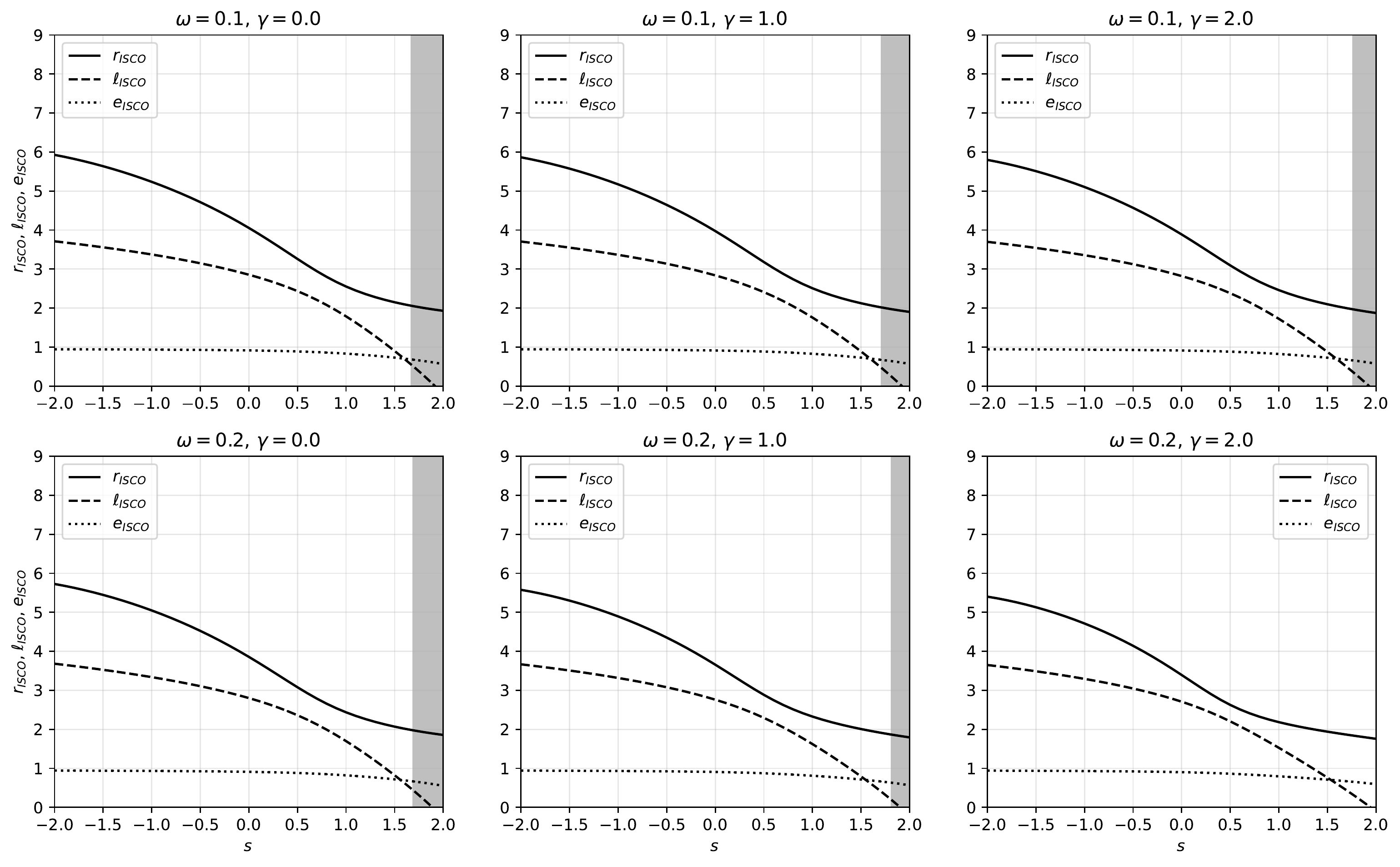}
    \caption{ISCO parameters as function of the spin $s$ for a test particle moving in the improved rotating black hole background with $a=0.5$. The shaded regions correspond to superluminal motion (using $G_0 = M =1$ and measuring $s$ in units of $m$).}
    \label{fig:quantum-ISCO}
\end{figure}

\begin{figure}[!htp]
    \centering
    \includegraphics[width=1\linewidth]{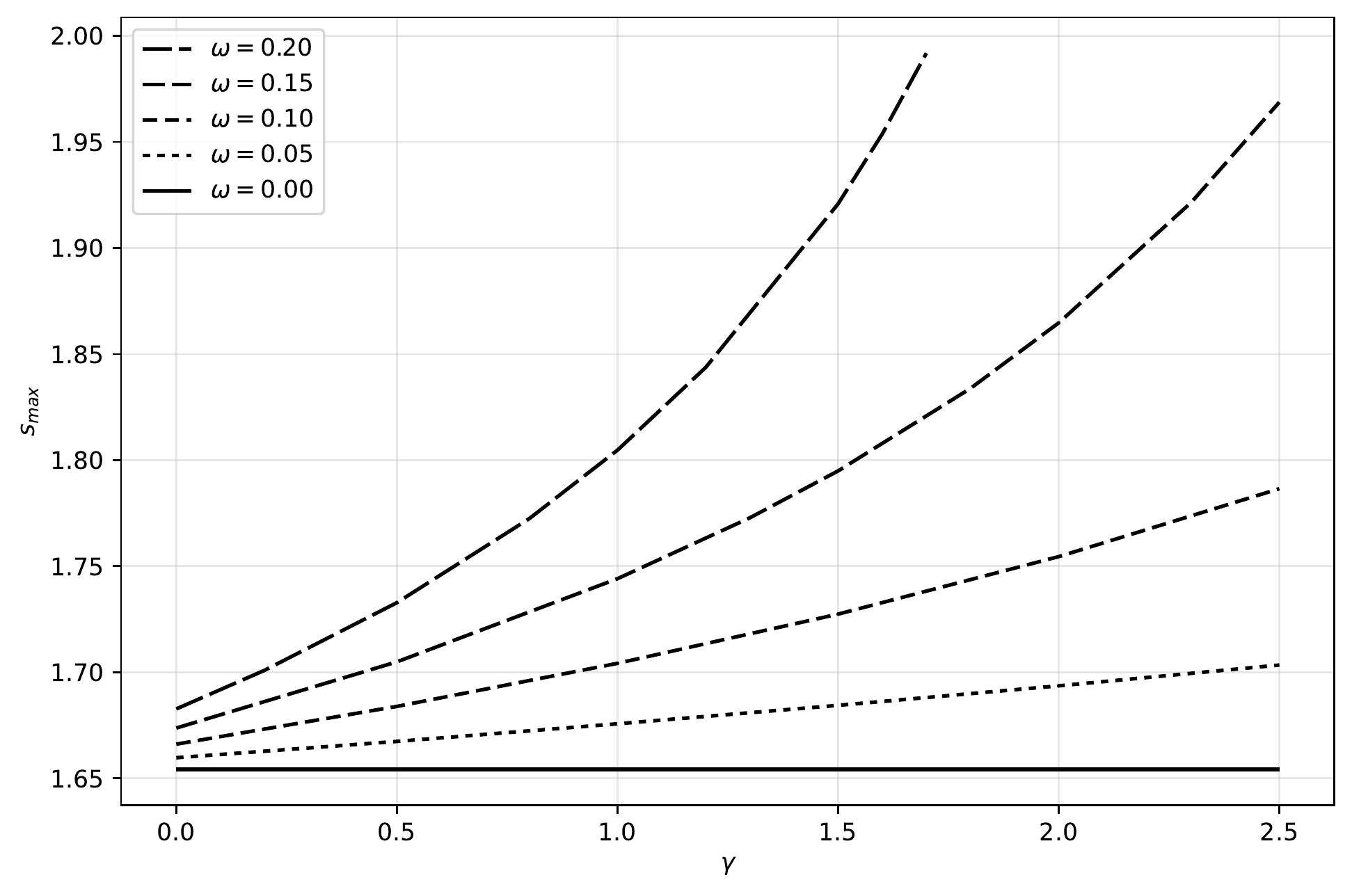}
    \caption{Maximum spin allowed by the velocity constrain as function of the parameters $\omega$ and $\gamma$ for the improved rotating black hole metric with $a=0.5$ (using $G_0 = M =1$ and measuring $s$ in units of $m$).}
    \label{fig:maxspin2}
\end{figure}

Similar results are obtained for the properties of the ISCO in the  quantum  improved  RBH background. As shown in Figure \ref{fig:quantum-ISCO}, the values of $r_{ISCO}$, $\ell_{ISCO}$ and $e_{ISCO}$ also decrease when the spin or the parameters $ \omega $ and $\gamma$ increase and it is clear that the modification caused by the quantum parameters is more noticeable for $r_{ISCO}$. Once more, non-physical motion is identified by the shaded regions in Figure \ref{fig:quantum-ISCO} and the values of $s_{max}$  as function of the quantum parameters is depicted in Figure \ref{fig:maxspin2}. 

All the previous results obtained numerically show that the ISCO parameters for a spinning test particle moving in the quantum improved RBH background are smaller than in any of the other cases studied in this work. Depending on the values of the quantum parameters $ \omega $ and $\gamma$, the larger they were taken, the smaller the ISCO parameters were and the larger the values of the maximum spin.

\section{Conclusion}

In this work, we have studied the motion of a spinning test particle in a quantum improved RBH background which introduces a position-dependent gravitational constant, $ G (r) $, and involves two quantum parameters, $ \omega $ and $\gamma$. By writing the equations of motion, the effective potential is determined and from it, we calculate the following ISCO parameters: the radius $r_{ISCO}$, the angular momentum $l_{ISCO}$ and the energy $e_{ISCO}$ as a function of the spin $s$ of the test particle.

As a first result, the ISCO parameters for the classical black holes ($ \omega  =0$), specifically both Schwarzschild ($a=0$) and  Kerr ($a \neq 0$) solutions, agree with the previously reported results in Refs. \refcite{Zhang2018,Jefremov2015}.

When considering the quantum improved Schwarzschild black hole background ($ \omega  \neq 0$ and $a=0$), our results show that the ISCO parameters decrease under an increase of the spin $s$ or under an increase of the parameters $ \omega $ and $\gamma$. Additionally, we have obtained the values of the maximum spin allowed by the condition of physical motion (timelike velocity) and its behavior shows that high values of the quantum parameters $ \omega $ and $\gamma$ imply a larger value of this maximum spin.

Similar results are obtained for the ISCO properties in the  quantum improved  RBH background. Once again, an increase in the quantum parameters $ \omega $ and $\gamma$ produce a decrease in the values found of $r_{ISCO}$, $l_{ISCO}$ and $e_{ISCO}$. In general, the introduction of the black hole spin parameter $a\neq 0$ implies a set of ISCO parameters that are smaller than those in the non-rotating case. 

Finally, from our analysis of the motion of a spinning test particle moving in the quantum improved non-rotating and rotating backgrounds it is notable that the effects caused by the introduction of the position-dependent gravitational constant from the QEG approach on the ISCO properties are similar to those obtained in other black holes backgrounds. Particularly, the results obtained for a spinning test particle moving in the quantum improved non-rotating and RBH backgrounds are similar to those already reported in Refs. \refcite{Zhang2018,Conde2019,Larranaga2020,Zhang2020} although in these cases the parameters that affect the properties of the ISCO come from different theories. For example, in Ref. \refcite{Zhang2018,Conde2019} the parameter of the electric charge $Q$ of the black hole is the one that affects the trajectories. In Ref. \refcite{Zhang2020}, the Gauss–Bonnet
coupling parameter $\alpha$ of the Einstein–Gauss–Bonnet 4D black hole is the one that affects the motion while, in Ref. \refcite{Larranaga2020} the  length scale in the Hayward black hole is the one changing the orbits. Therefore, the results for the quantum improved non-rotating and RBH backgrounds would compete with the modifications in these cases from an observational point of view.

\section*{Acknowledgments}

This work was supported by the Universidad
Nacional de Colombia. Hermes Grant Code 55070.

\appendix

\section{Appendices}

In this appendix, we present the inverse of the metric tensor in \eqref{eq:Metric},
\begin{equation}
\begin{aligned}
g^{tt}&= -\dfrac{(r^2 + a^2)^2 - a^2\Delta \sin^2 \theta}{\Delta \Sigma} \\
g^{t\phi} &= - \dfrac{2MG(r) a r}{\Delta \Sigma}\\
g^{rr} &= \dfrac{\Delta}{\Sigma} \ \ \ \ \ \  \ g^{\theta \theta} = \dfrac{1}{\Sigma} \\
g^{\phi \phi} &= \dfrac{\Delta - a^2 \sin^2\theta}{\Delta \Sigma \sin^2 \theta},
\end{aligned}
\label{eq:inverse}
\end{equation}
needed in the geodesic calculation. We also include here the relevant non-vanishing Riemann tensor components appearing in the MPD equation for the spinning test particle,

\begin{align}
R_{3003} = &\frac{M r \sin ^2\theta  \left(a^2-2 M r G(r)+r^2\right) }{\left(a^2 \cos ^2\theta +r^2\right)^3} \left[r G'(r) \left(a^2 \cos ^2\theta +r^2\right)\right.\notag \\
& \left.-G(r) \left(r^2-3
   a^2 \cos ^2\theta \right)\right],
\end{align}

\begin{align}
R_{3113} = & -\frac{M \sin ^2\theta }{4 \left(a^2-2 M r G(r)+r^2\right) \left(a^2 \cos ^2\theta +r^2\right)^3}\Bigl\{-a^2 r \sin ^2\theta  G''(r) \left(a^2 \right.\notag \\
& \left. -2 M r G(r)+r^2\right) \left(a^2 \cos
   (2 \theta )  +a^2+2 r^2\right)^2 -8 a^2 M r^2 G(r)^2 \sin ^2\theta \notag \\
&  \left(3 a^2 \cos (2 \theta )+3 a^2-2 r^2\right)-4 r
   \left(a^2+r^2\right) G(r) \left(r^2
    -3 a^2 \cos ^2\theta \right)  \notag \\
   & \left(-a^2 \cos (2 \theta ) +2 a^2+r^2\right)+G'(r) \left[2
   a^2 M r G(r) \sin ^2\theta  \left(a^4 (4 \cos (2 \theta )\right. \right.    \notag \\
   & \left.+\cos (4 \theta ))+3 a^4-8 r^4\right)  +\left(a^2+r^2\right)
   \left(a^2 \cos ^2\theta +r^2\right)  \left(a^4 \cos (4 \theta )\right.    \notag \\
   & \left. \left.-a^4-4 a^2 r^2 \cos (2 \theta )+8 a^2 r^2+4
   r^4\right)\right]\Bigr\},
\end{align}

\begin{align}
R_{3101} = &\frac{a M \sin ^2\theta  }{\left(a^2-2 M r G(r)+r^2\right) \left(a^2 \cos
   ^2\theta +r^2\right)^3}\Bigl\{\frac{1}{4} r G''(r) \left(a^2-2 M r G(r) +r^2\right) \notag \\
   &\left(a^2 \cos (2
   \theta )+a^2+2 r^2\right)^2-4 M r^2 G(r)^2 \left(r^2-3 a^2 \cos ^2\theta \right)+3 r \notag \\
   & \left(a^2+r^2\right) G(r) \left(r^2-3
   a^2 \cos ^2\theta \right)+G'(r) \left[4 M r G(r) \left(r^4-a^4 \cos ^4\theta \right)\right.\notag \\
   & \left.  +\left(a^2+r^2\right) \left(a^2 \cos
   ^2\theta +r^2\right) \left(2 a^2 \cos ^2\theta -3 r^2\right)\right]\Bigr\},
\end{align}

\begin{align}
R_{1001} = & \frac{M}{\left(a^2-2 M r G(r)+r^2\right)
   \left(a^2 \cos ^2\theta +r^2\right)^3}\Bigl\{\frac{1}{4} r G''(r) \left(a^2-2 M r G(r)+r^2\right)\notag \\
   & \left(a^2 \cos (2 \theta )+a^2+2 r^2\right)^2-4 M r^2 G(r)^2
   \left(r^2-3 a^2 \cos ^2\theta \right)+r G(r) \notag \\
   & \left(r^2-3 a^2 \cos ^2\theta \right) \left(-a^2 \cos ^2\theta
   +3 a^2+2 r^2\right)+G'(r) \left[4 M r G(r)\right.\notag \\
     & \left(r^4
     -a^4 \cos ^4\theta \right)-\left(a^2 \cos ^2\theta +r^2\right)
   \left(-a^2 \left(2 a^2+3 r^2\right) \cos ^2\theta  \right. \notag \\
   & \left. \left. +3 a^2 r^2+2 r^4\right)\right]\Bigr\},
\end{align}

where

\begin{align}
G'(r) = \frac{r^2  \omega   G_o^2 \left(3 \gamma  M G_o+2 r\right)}{\left( \omega   G_o \left(\gamma  M G_o+r\right)+r^3\right){}^2} \label{eq:firstderG}
\end{align}

and

\begin{align}
G''(r)= \frac{2 r  \omega   G_o^2 \left[G_o \left(3 \gamma  M  \omega   G_o \left(\gamma  M G_o+r\right)+r^2 ( \omega  -6 \gamma  M
   r)\right)-3 r^4\right]}{\left( \omega   G_o \left(\gamma  M G_o+r\right)+r^3\right){}^3}.
\end{align}

%\begin{thebibliography}{000} %for 3 digits
%\begin{thebibliography}{00}  %for 2 digits

\end{document}